\documentclass[aps,prl,twocolumn,showpacs,groupedaddress]{revtex4-1}  
\usepackage{graphicx}  
\usepackage{dcolumn}   
\usepackage{bm}        
\usepackage{amssymb}   

\usepackage{epstopdf}

\begin{document}

\title{Hadronic $B_c$ decays as a test of $B_c$ cross-section}
\author{Alexander Rakitin}
\address{Lauritsen Laboratory, California Institute of Technology, Pasadena, California 91125} 
\author{Sergey Koshkarev}
\address{Instituto de F\'isica de Cantabria, CSIC-University of Cantabria, Santander, Cantabria E-39005, Spain}

\date{\today}

\begin{abstract}
  This paper focuses on disagreement between theoretical predictions
  and experimental results of the production properties of $B_c$
  meson. Hadronic decays of $B_c$ are used to separate predictions of
  production cross-section and predictions of branching ratio. The
  branching ratios of $B_c$ decays to $J/\psi + \pi$ and to $J/\psi + 3
  \pi$ are also presented.
\end{abstract}

\maketitle 

\section{Introduction}
\label{sec:Intro}

\begin{figure}[h]
\includegraphics[scale=0.165]{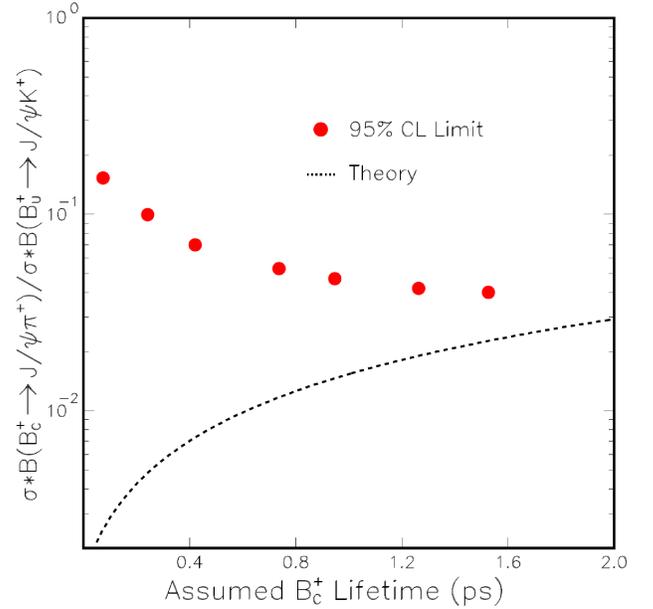}
\caption{\label{fig:BcRunI} The circular points show the different
  95\% C.L. on the ratio of cross-section times branching fraction for
  $B_{c}^{+} \to J/\psi \pi^{+}$ relative to $B_{u}^{+} \to J/\psi
  K^{+}$ as a function of the $B_{c}^{+}$ lifetime. The dotted curve
  represents calculation of this ratio based on the assumption that
  the $B_{c}^{+}$ is produced $1.5 \times 10^{-3}$ times as often as
  all other $B$ mesons and that $\Gamma(B_{c}^{+} \to J/\psi \pi^{+})
  = 4.2 \times 10^{9} ~s^{-1}$.}
\end{figure}

Study of $B_c$ meson is important because it stands out of the crowd
of other heavy-quark mesons. This is the only meson consisting of two
different heavy quarks. Also, the lighter $c$ quark has a decay rate
($\sim$65\%)~\cite{Kiselev2003} larger than heavier $b$ quark, which
is uncommon for heavy-quark mesons. The mass and lifetime of $B_c$
meson have been measured by CDF~\cite{CDFmass, CDFlifetime} and
D\O~\cite{d0mass, d0lifetime} in decays $B_c\to J/\psi\pi$ and $B_c\to
J/\psi \ell$. They are in pretty good agreement with theory
~\cite{Kiselev2003, TWQCD} (see Table~\ref{tab:table1}).
\begin{table*}
\begin{tabular}{|l|c|c|}
\hline
Source&$B_{c}$ mass $({\rm MeV}/c^{2})$&$B_{c}$ lifetime $({\rm ps})$\\
\hline
CDF~\cite{CDFmass, CDFlifetime} & $6285 \pm 5.3(stat) \pm 1.2(sys)$ &$0.463^{+0.073}_{-0.65}(stat) \pm 0.036(sys)$\\
D\O~\cite{d0mass, d0lifetime} & $6300 \pm 14(stat)\pm 5(sys)$ & $0.448^{+0.038}_{-0.036}(stat) \pm 0.032(sys)$ \\
Theory~\cite{Kiselev2003,TWQCD} & $6278$ & $0.48\pm0.05$ \\
\hline
\end{tabular}
  \caption{\label{tab:table1}
    This table shows good agreement of theoretical predictions of $B_c$ properties with experimental results from Tevatron.
  }
\end{table*}
Also, the production
properties of $B_c$ meson have been measured and compared to that of
$B$ meson~\cite{Vaia2005}:
\[
R_{e} = \frac
{\sigma(B_{c}) \cdot Br(B_{c} \rightarrow J/\psi e^{\pm} \nu)}
{\sigma(B) \cdot Br(B \rightarrow J/\psi K^{\pm})} 
= 0.282 \pm 0.038 \pm 0.074
\]
and
\[
  R_{\mu} = \frac
{\sigma(B_{c}) \cdot Br(B_{c} \rightarrow J/\psi \mu^{\pm} \nu)}
{\sigma(B) \cdot Br(B \rightarrow J/\psi K^{\pm})} 
= 0.249 \pm 0.045_{-0.076}^{+0.107}
\]
in the kinematic region $p_{T}(B_{(c)}) > 4.0 ~{\rm GeV}$ and
$|y(B_{(c)})| < 1.0$. Using the theoretical predictions for the
branching fraction $Br(B_{c} \rightarrow J/\psi e^{+} \nu) \approx 2
\cdot 10^{-2}$~\cite{Kiselev2003, Sanchis1995} and taking into account
well-measured branching $Br(B^{+} \rightarrow J/\psi K^{+}) =
(1.007\pm0.0035)\cdot 10^{-3} $~\cite{PDG}, one can obtain the ratio
of the production cross-sections:
\[
\frac{\sigma(B_{c})}{\sigma(B)} 
= R_{e} \cdot \frac
{Br(B \rightarrow J/\psi K^{\pm})}
{Br(B_{c} \rightarrow J/\psi e^{\pm} \nu)} 
\simeq 1.4 \cdot 10^{-2}.
\]
Comparing this result with theoretical predictions of $B_{c}$
cross-section ~\cite{Bereznoy1996, Kolodziej1998, FermilabReview2001,
  Braaten} and of the ratio of production cross-section $\sim 10^{-3}$
we see that $B_{c}$ semileptonic branching fraction has to be an order
of magnitude larger than theoretical prediction, about 20\%. This is a
significant discrepancy between theory and experiment.  Another
discrepancy comes from the measurement of the production properties of
$B_{c}$ in CDF data collected in Run\,I~\cite{RunI}. CDF presented a
95\% C.L. on $\sigma(B_{c}^{+}) \cdot Br(B_{c}^{+} \rightarrow J/\psi
\pi^{+}) / \sigma(B_{u}^{+}) \cdot Br(B_{u}^{+} \rightarrow J/\psi
K^{+})$ as a function of $B_c$ lifetime (see Fig.~\ref{fig:BcRunI}).
Using known $B_c$ lifetime $(0.46\pm0.07) ~ps$ we clearly see an order
of magnitude disagreement between the theoretical prediction and
data. Either our theoretical estimate of $B_c$ semileptonic branching
fraction is incorrect or we do not understand the production
cross-section of $B_c$.  To clarify this issue we suggest to measure
the ratio of the production cross-sections using hadronic decay modes
of $B_c$, namely $B_c\to J/\psi \pi$ and $B_c\to J/\psi 3\pi$. If the
experimental branching fraction ratio coincides with theoretical
predictions, the problem is in the production cross-section,
otherwise the prediction for $B_c$ semileptonic branching fractions is
incorrect. This measurement can be done in CDF or D\O, where $B_c$
mesons are produced and already were observed.
%
%
%
\section{Theoretical basement}
In this paper we will use the fact that a hadronic matrix element of
heavy-quark current might be written in a simple form if expressed in
terms of the velocities of heavy particles~\cite{Sanchis1995,
  Neubert1991, Isgur1989}.  Also, we will base on definition of
nonrecoil form factor. The validity of using the nonrecoil
approximation is strongly supported by the fact that the kinematic
variable $\omega = \upsilon_{1} \cdot \upsilon_{2}$ is restricted to
values close to unity (indexes 1 and 2 do mean initial and final
hadrons, respectively). Let heavy-quark $Q_{i}$ undergo a weak decay
to $Q_{f}$ with a spectator quark $Q_{s}$.  At $s = (p_{1} - p_{2})^2
= 0$, we have
\begin{eqnarray}
	\nonumber
	(\upsilon_{1}\cdot\upsilon_{2})_{max} = 1+ \frac{(m_{1} - m_{2})^{2}}{2m_{1}m_{2}}\\
	 \simeq 1+ \frac{(m_{Q_{i}} - m_{Q_{f}})^{2}}{2(m_{Q_{i}} + m_{Q_{f}})(m_{Q_{}} + m_{Q_{s}})}.
\end{eqnarray}
In our case the initial state $(bc)$ decays into $(cc)$, therefore $(\upsilon_{B_{c}} \cdot \upsilon_{J/\psi})_{max} \simeq$ 1.29
for the mass values $m_{b} = 4.8~{\rm GeV}/c^{2}$ and $m_{c} = 1.5~{\rm GeV}/c^{2}$.

Following~\cite{Sanchis1995} the hadronic matrix elements in the nonrecoil approximation $(\upsilon_{i} = \upsilon_{f} = \upsilon)$ for the weak process $Q_{i} \rightarrow Q_{f} W^{*}$ can be presented as
\begin{eqnarray}
\nonumber
	<0^{P}, \epsilon_{2} | V^{\mu} | 0^{P}>  & \simeq & \pm 2 \eta_{12} \sqrt{m_{1}m_{2}} \upsilon^{\mu} \\
	\nonumber
	<1^{P}, \epsilon_{2} | A^{\mu} | 0^{P}>  & \simeq & \pm 2 \eta_{12} \sqrt{m_{1}m_{2}} \epsilon_{2}^{*\mu} \\
	\label{eq:aaa}
	<0^{P}| A^{\mu} | 1^{P}, \epsilon_{1}> & \simeq & \pm 2 \eta_{12} \sqrt{m_{1}m_{2}} \epsilon_{1}^{\mu} ,\\
	\nonumber
	< 1^{P}, \epsilon_{2} | V^{\mu} | 1^{P}, \epsilon_{1} > & \simeq & \pm 2 \eta_{12} \sqrt{m_{1}m_{2}} (\epsilon_{1} \cdot \epsilon_{2}^{*}) \upsilon^{\mu} \\
	\nonumber
	< 1^{P}, \epsilon_{2} | A^{\mu} | 1^{P}, \epsilon_{1} > & \simeq & \pm 2 \eta_{12} \sqrt{m_{1}m_{2}} i \varepsilon^{\mu \nu \alpha \beta} \upsilon_{1\nu} \epsilon_{1 \alpha} \epsilon_{2 \beta}^{*}
\end{eqnarray}
where the vector and axial-vector currents are $V^{\mu} = \bar{Q_{f}}
\gamma^{\mu} Q_{i}$ and $A^{\mu} = \bar{Q_{f}} \gamma^{\mu} \gamma_{5}
Q_{i}$ and $\eta_{12}$ is a form factor playing the role of Isgur-Wise
functions for transition between initial and final states of
hadrons. Here, $\eta_{12}$ can be parametrized as
\[
	\eta_{12}  = (\frac{2\beta_{1}\beta_{2}}{\beta_{1}^{2} + \beta_{2}^{2}})^{3/2}.
\]
For the case of $B_{c}$ decays to $J/\psi$, the parameters $\beta_{1}$ and $\beta_{2}$ are equal to 0.82 and 0.66, respectively.

\section{decay $B_{c}^{+}$ to $J/\psi + \pi^{+}$}
\label{sec:JPsiPi}

The amplitude of this decay includes two factors, one of them is a
pionic decay amplitude, and the other is the formfactor appearing in
semileptonic decay. This gives us a direct relation between pionic
and semileptonic decays. In the case $s = m_{\pi}^{2} \simeq 0$, the width of
pionic decay may be given as~\cite{Bjorken1988}
\begin{equation}
  \frac{\Gamma (B_{c} \rightarrow J/\psi + \pi)}{d\Gamma /ds|_{s=0}(B_{c} \rightarrow J/\psi + \bar{\ell}\nu)} \simeq 6 \pi^{2} f_{\pi}^{2}|V_{ud}|^{2} \simeq 1~{\rm GeV^2}.
	\label{eq:bbb}
\end{equation}
Upon contracting Eq.~\ref{eq:aaa} with leptonic current $\bar{\ell}
\gamma^{\mu}(1-\gamma_{5})\nu$, the width of $B_{c} \rightarrow J/\psi +
\bar{\ell}\nu$ ~\cite{Sanchis1995} is

\begin{equation}
	\frac{d\Gamma}{ds} \simeq 3 \cdot \frac{G_{F}^{2}(\lambda^{3/2} + 12sm_{B_{c}}^{2}\lambda^{1/2})}{576 \pi^{3} m_{B_{c}}^{3}} \frac{m_{J/\psi}}{m_{B_{c}}} \eta_{B_{c}J/\psi}^{2}  |V_{cb}|^{2}, 
	\label{eq:ccc}
\end{equation}
 where $\lambda \equiv \lambda(m_{J/\psi}^{2},m_{B_{c}}^{2},s) $ is "triangle" K\"allen function denoted as
 \[
 	\lambda(x,y,z) = (x^{2} + y^{2} + z^{2} - 2xy - 2yz -2zx)^{1/2}.
\]
Combining Eqs. \ref{eq:bbb} and \ref{eq:ccc} and using $B_c$ lifetime
$\tau = 0.46~ps$, we may expect the pionic decay branching ratio to
be
 
\[
  Br(B_{c}^{+} \rightarrow J/\psi + \pi^{+}) \simeq 0.2\%.
\]

This result is in good agreement with other results (see
~\cite{Kiselev2003, Sanchis1995} and references therein).

\section{Decay $B_{c}^{+}$ to $J/\psi + \pi^{+} \pi^{-} \pi^{+}$}
\subsection{Axial current}
The amplitude of $B_{c}^{+} \rightarrow J/\psi + \pi^{+} \pi^{-} \pi^{+}$ is
\begin{equation}
	A \sim < J/\psi | A^{\mu} |B_{c} >  <\pi^{+} \pi^{-} \pi^{+} |J_{axial}^{\mu}(0) | 0>,
\end{equation}
where $ < J/\psi | A^{\mu} |B_{c} > =  2 \eta_{B_{c}J/\psi} \sqrt{m_{B_{c}}m_{J/\psi}} \epsilon_{J/\psi}^{*\mu}$, $A^{\mu} = \bar{c} \gamma^{\mu} \gamma_{5} b$ is the axial-vector current and  $\epsilon_{J/\psi}^{*\mu}$ presents the polarization four-vector of $J/\psi$.
Let us remind the reader that the phase space can be represented as
\begin{eqnarray*}
	dPS(B_{c} \rightarrow J/\psi + 3\pi) & \\	
	 = \frac{ds}{2\pi} dPS & (B_{c} \rightarrow J/\psi W^{*}) dPS(W^{*} \rightarrow 3\pi), 
\end{eqnarray*}
where the three-pion phase space is
\begin{eqnarray}
\nonumber
	\frac{1}{2\pi} \int dPS (3\pi) <0|J_{\mu}^{\dagger}|3\pi> <3\pi|J_{\nu}|0> \\
	= q_{\mu}q_{\nu}\rho_{0}(s) + (q_{\mu}q_{\nu} - g_{\mu\nu}q^{2})\rho_{1}(s),
\end{eqnarray}
where $q$ the four momentum vector of the three-pion state and s = $q^{2}$. It is easy to show that $\rho_{0} = 0$.

We have the same situation in $\tau^{-} \rightarrow \nu_{\tau} +
\pi^{-} \pi^{+} \pi^{-}$ decay, therefore we will follow $a_{1}$ meson
domination model of Ref.~\cite{Kuhn1990} ($a_{1}$ dominance is
also discussed in ~\cite{Li1997, Ivanov1989}, angular distributions of
$\tau \to \nu + 3\pi$ are discussed in ~\cite{Kuhn1992}).  The spectral
function $\rho_{1}(s)$ can be cast into the form:

\begin{equation}
	\rho_{1}(s) = \frac{1}{6} \frac{1}{(4\pi)^{4}} \frac{8}{9f_{\pi}^{2}} |BW_{a_{1}}(s)|^{2} \frac{g(s)}{s}.
\end{equation}

The Breit-Wigner function $BW_{a_{1}}$ is parametrized including energy dependent width $\Gamma_{a_{1}}(s)$:
\begin{equation}
	BW_{a_{1}} = \frac{m_{a_{1}}^{2}}{m_{a_{1}}^2 - s - i\sqrt{s}\Gamma_{a_{1}}(s)},~\Gamma_{a_{1}}(s) = \frac{m_{a_{1}}}{\sqrt{s}} \Gamma_{a_{1}} \frac{g(s)}{g(m_{a_{1}}^{2})}
\end{equation}
where $m_{a_{1}} = 1251\pm13~{\rm GeV}$, $\Gamma_{a_{1}} = 599 \pm
44~{\rm MeV}$, and the function $g(s)$ has been calculated in
Ref.~\cite{Kuhn1990} and is derived from the observation that the
axial-vector resonance $a_{1}$ decays predominately into tree
pions. In this way, the branching is
\[
	Br(B_{c}^{+} \rightarrow J/\psi + \pi^{+} \pi^{-} \pi^{+}) \simeq 0.3 \%.
\]
\subsection{Vector current}
The other possibility to observe three charged pions in fully
reconstructed mode is $B_{c}$ decay to $J/\psi + \omega \pi$, where
$\omega$ decays to $\pi^{+}\pi^{-}$. However, the simple analysis of
similar decay $\tau^{-} \to \nu_{\tau}  + \omega \pi^{-}, \omega \to \pi^+\pi^-$ decay 
shows that this mode gives a too small contribution.
\section{Summary}

Current theoretical and experimental knowldege about $B_c$ meson
suggests that either we do not understand the production cross-section
or semileptonic branching fraction of $B_c$ (see
Sec.
1). In our paper we propose to measure the
branching fractions for $B_c$ decays into final states $J/\psi + \pi$
and $J/\psi + 3\pi$ to resolve this issue. Since the decays to
$J/\psi + \pi$ and $J/\psi + \ell \nu$ are correlated (as discussed in
Sec.
3) the decay into $J/\psi + 3\pi$ has a special
meaning, allowing for independent test of $B_c$ production
cross-section. The predictions of the branching fractions of $B_c$
decays into these final states are also obtained.
\acknowledgements{
We would like to acknowledge Maiko Takahashi's incredible involuntary spiritual support.
We would also like to thank Alberto Ruiz for much useful advice.}

\end{document}